\documentclass{aa}  

\usepackage{float}
\usepackage{graphicx}
\usepackage{txfonts}
\usepackage[table]{xcolor}
\usepackage{natbib}
\bibpunct{(}{)}{,}{a}{}{,}  
\usepackage{pdflscape}
\usepackage{hyperref}
\hypersetup{breaklinks=true,
            colorlinks,
            filecolor=blue,
            linkcolor=blue,
            citecolor=blue,
            urlcolor=blue,
            anchorcolor=blue} 
\usepackage{listings}
\usepackage{xcolor}

\definecolor{codegreen}{rgb}{0,0.6,0}
\definecolor{codegray}{rgb}{0.5,0.5,0.5}
\definecolor{codepurple}{rgb}{0.58,0,0.82}
\definecolor{backcolour}{rgb}{0.95,0.95,0.92}

\lstdefinestyle{mystyle}{
    backgroundcolor=\color{backcolour},   
    commentstyle=\color{codegreen},
    keywordstyle=\color{magenta},
    numberstyle=\tiny\color{codegray},
    stringstyle=\color{codepurple},
    basicstyle=\ttfamily\footnotesize,
    breakatwhitespace=false,         
    breaklines=true,                 
    captionpos=b,                    
    keepspaces=true,                 
    numbers=left,                    
    numbersep=5pt,                  
    showspaces=false,                
    showstringspaces=false,
    showtabs=false,                  
    tabsize=2
}

\begin{document}

   \title{\texttt{pastamarkers} 2: pasta sauce colormaps for your flavorful results}

   \author{PASTA Collaboration
          \and
          L. Rosignoli
          \inst{1,2}            
          \and
          A. Della Croce
          \inst{2}
          \and
          E. Leitinger
          \inst{1}
          \and
          L. Leuzzi
          \inst{2}
          \and   
          G. Papini 
          \inst{1, 2} 
          \and           
          A. Traina
          \inst{2} 
          \and
          S. Sartori
          \inst{3}
          \and
          N. Borghi
          \inst{1,2}
          \and   
          E. Ceccarelli
          \inst{1,2}
          }

   \institute{
            Department of Physics and Astronomy, University of Bologna, Via Gobetti 93/2, 40129 Bologna, Italy
        \and
            INAF - Astrophysics and Space Science Observatory of Bologna, Via Gobetti 93/3, 40129 Bologna, Italy 
        \and
            Aix-Marseille Universit\'e, CNRS/IN2P3, CPPM, Avenue De Luminy 163, 13288 Marseille, France
            }

   \date{Received xxxx; accepted yyyy}

 
  \abstract
   {}
   {In the big data era of Astrophysics, the improvement of visualization techniques can greatly enhance the ability to identify and interpret key features in complex datasets. This aspect of data analysis will become even more relevant in the near future, with the expected growth of data volumes. With our studies, we aim to drive progress in this field and inspire further research.}
   {We present the second release of \texttt{pastamarkers}, a \texttt{Python}-based \texttt{matplotlib} package that we initially presented last year \citep{patamarkers_2024}. In this new release we focus on big data visualization and update the content of our first release.}
   {We find that analyzing complex problems and mining large data sets becomes significantly more intuitive and engaging when using the familiar and appetizing colors of pasta sauces instead of traditional colormaps.}
   {}


   \maketitle
%
\section{Introduction}\label{sec:intro}
With the recent launch of large photometric surveys and the imminent start of others, Astrophysics is entering the big data era, as many other scientific fields. As a result, the volume of data that astrophysicists must process is bound to increase dramatically, creating an urgent need to develop and test new pre-analysis and analysis methods to effectively manage this transition. This is needed to reduce the workload of researchers and computational resources. 

Compared to other sciences, astronomical data are inherently complex, often characterized by hidden patterns and unforeseen relations that in some cases are still poorly understood. Before applying robust statistical methods to study the properties of such data sets, it is crucial for researchers to develop an intuitive understanding of their structure. This makes effective visualization tools highly important, as they provide a first, essential step in exploring the data, identifying trends, and guiding subsequent analysis with greater insight.

In our previous work \citep{patamarkers_2024}, we have studied the effectiveness of using pasta-shaped markers for the visualization of data in different branches of Astrophysics. Given the good response and  growing interest from the community, we have expanded our research and present new release our \texttt{pastamarkers} in this work.

This paper is organized as follows: in Sect. \ref{sec:pastamarker} we revise the content of the first release and present the second release; in Sect. \ref{sec:examples} we show a few examples of the usage of the colormaps we peopose; we conclude in Sect. \ref{sec:conclusions}. 

\section{\texttt{pastamarkers 2}}\label{sec:pastamarker}

\subsection{The first release}
In \cite{patamarkers_2024}, we have released the \texttt{pastamarkers} package, a publicly available open source code\footnote{You can download it at \url{https://github.com/LR-inaf/pasta-marker}.} for expanding the library of available markers of \texttt{matplotlib} with unconventional pasta-shaped markers. 

\subsection{New geometry for colored pasta}
\begin{figure*}[!th]
    \centering
    \includegraphics[width=\textwidth]{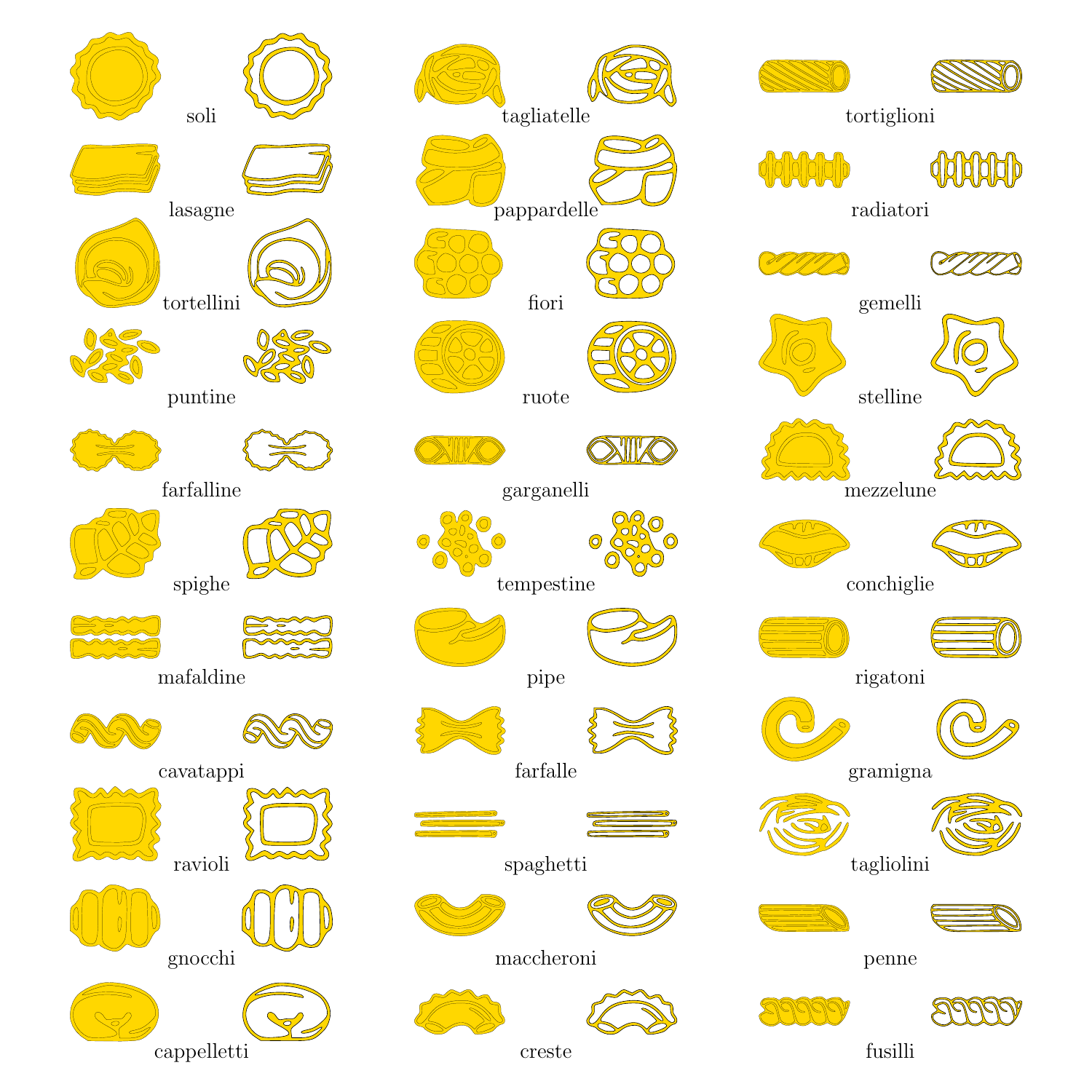}
    \caption{Showcase the different color scheme between the 2024 release and the one presented in this paper of the \texttt{pastamarkers} package. Yellow-filled pasta types are the second release of the markers, to be compared with the first one (empty markers) that could not be properly colored.}
    \label{fig:filled_pastas}
\end{figure*}

With this release of the \texttt{pastamarkers} package \citep{patamarkers_2024} we provide the users with an updated version of the 11 pasta types to be used in combination with pasta sauces (see Sect.~\ref{sec:colormaps}).
The new markers can now be properly colored through the usual \texttt{matplotlib} inputs such as \texttt{color} or \texttt{markerfacecolor}, while in the previous version, only the regions between the inner and outer edges of the marker were colored. Figure~\ref{fig:filled_pastas} shows the different outputs for the two versions of the \texttt{pastmarkers} package.

\subsection{Colormaps}\label{sec:colormaps}

\begin{figure}[!th]
    \centering
    \includegraphics[width=.49\textwidth]{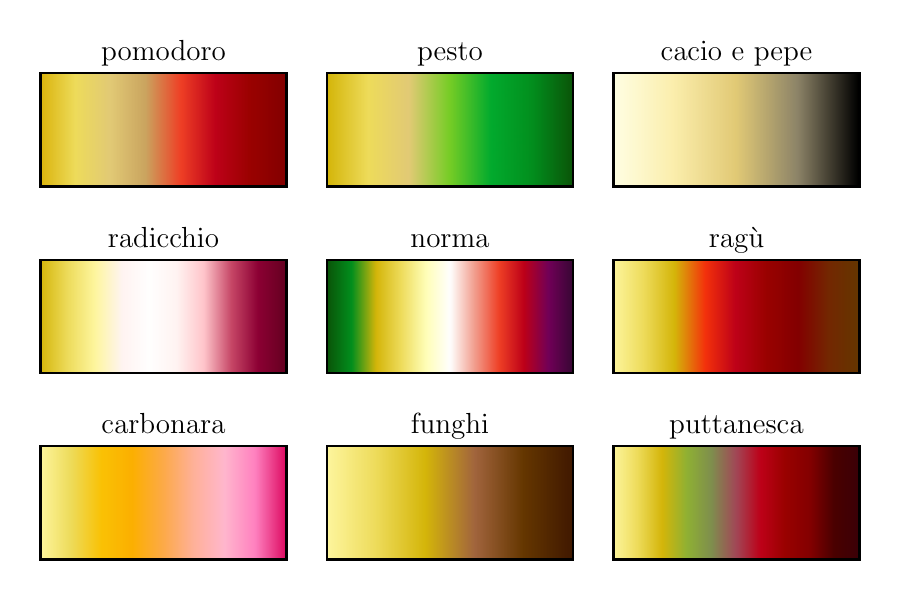}
    \caption{Showcase of the pasta sauce colormaps available in the new release of the \texttt{pastamarkers} package.}
    \label{fig:cmaps}
\end{figure}

With this release, we introduce nine colormaps to accompany your favorite pasta shapes, based on popular Italian pasta sauces. Each colormap includes specific colors tailored to mimic the ingredients contained within each sauce, which we describe in the sections below and showcase in Figure \ref{fig:cmaps}.\\ 
We also offer a version of some colormaps which end with a white topping of parmigiano, indicated by the addition of `\texttt{\_p}' to the original pasta sauce name. These parmigiano versions are available for all colormaps except: cacio e pepe, radicchio and pasta alla norma, as these colormaps already contain a white component. Alternatively, we also include a new function (\texttt{`add$\_$parmesan'}) designed to randomly scatter white points over the pasta markers and sauce, to mimic the dusting of parmigiano over the dish.\\

\subsubsection{Pomodoro (\texttt{salsa.pomodoro})}
An Italian staple and the base of many other sauces, `pomodoro' translates to tomato, so this simple tomato sauce colormap contains the iconic red of the tomatoes, beige for onion and garlic, and a golden yellow for the pasta itself.\\
\subsubsection{Pesto (\texttt{salsa.pesto})}
The vibrant green hues of this colormap correspond to the many shades of green you observe from crushing fresh basil. We include beige for both the pine nuts and garlic, with beige/yellow for the pasta.\\
\subsubsection{Cacio e pepe (\texttt{salsa.cacio\_e\_pepe})}
Simply translated, `cacio e pepe' is `cheese and pepper'. In this colormap we present a smooth gradient from the strong black of the peppercorns, through to the soft yellow of the pasta, with the creamy white colors symbolising the pecorino cheese.
\subsubsection{Radicchio (\texttt{salsa.radicchio})}
Radicchio sauce can be made a variety of ways, so we opted to keep things simple with a clean gradient representing the purple/red to white transition characteristic of radicchio leaves, before blending the white into the beige/yellow of the pasta.
\subsubsection{Pasta alla norma (\texttt{salsa.norma})}
Translated, the original name of `pasta alla norma' was simply `pasta with aubergine'; however, this pasta sauce has many ingredients which combine to make a truly colorful sauce. In this colormap, purple represents the star of the dish - aubergine - alongside the classic red of the tomato sauce, the white of the ricotta, the yellow of the pasta and finally the green of the basil. \textit{Note}: not color-blind friendly.
\subsubsection{Rag\`u (\texttt{salsa.ragu})}
Another classic Italian sauce, the ragu colormap contains brown for the meat, a classic red tomato sauce, orange for the carrots and yellow/beige for the pasta, onion and garlic.
\subsubsection{Carbonara (\texttt{salsa.carbonara})}
The carbonara colormap is bold and bright, with a range of pink hues for the guanciale (pork cheek in English), orange representing the egg yolks and yellow and beige for the pasta and pecorino cheese.
\subsubsection{Funghi (\texttt{salsa.funghi})}
Many Italian sauces use mushrooms as the base of the sauce, but with minor variations to the accompanying ingredients to produce a wide variety of `funghi' sauces. In this colormap we keep only a range of brown hues to symbolize the mushrooms, with also the beige/yellow of the pasta.
\subsubsection{Puttanesca (\texttt{salsa.puttanesca})}
Originating in Naples, Puttanesca literally translates to anchovies, olives and capers. We use very dark brown hues to symbolize the olives, the classic red of the tomato sauce, a slight purple hue for the anchovies, green for the capers and beige/yellow for the pasta. \textit{Note}: not color-blind friendly.\\

\subsection{Package usage}

The \texttt{pastamarkers} is part of Pypi repositories, so to install it simply type \texttt{pip install pastamarkers} on your python environment.
For development purposes, you can find the Github repo at \href{https://github.com/LR-inaf/pasta-marker}{https://github.com/LR-inaf/pasta-marker}.\\
After the package has been installed you can exploit the full potential of this module with a few lines of code.
\begin{lstlisting}[language=python]
from pastamarkers import pasta, salsa, add_parmesan
\end{lstlisting}
where in \texttt{pasta} you will find all the {\it pastamarkers} and in {\it salsa} the new colormaps.
In addition a new functionality has been implemented that allows you to \texttt{add\_parmesan} in every plot you want. This function takes the \texttt{matplotlib.pyplot.Axes} of your code and adds the parmesan on top of it. to use it just call it with the axes as input like the following example:
\begin{lstlisting}[language=python]
ax_with_parmesan = add_parmesan(ax)
\end{lstlisting}
finally we give here an end-to-end example of usage:

\begin{lstlisting}[language=python, caption=Python example]
import numpy as np
from pastamarkers import pasta, salsa, add_parmesan
x = np.random.uniform(-1, 1, 1000)
y = np.random.uniform(-1, 1, 1000)
z = np.random.uniform(-1, 1, 1000)
fig, ax = plt.subplots()
ax.scatter(x, y, marker=pasta.farfalle, c=z, cmap=salsa.pesto)
ax_with_parmesan = add_parmesan(ax)
\end{lstlisting}

\section{Examples}\label{sec:examples}

\begin{figure}[!th]
    \centering
    \includegraphics[width=.50\textwidth]{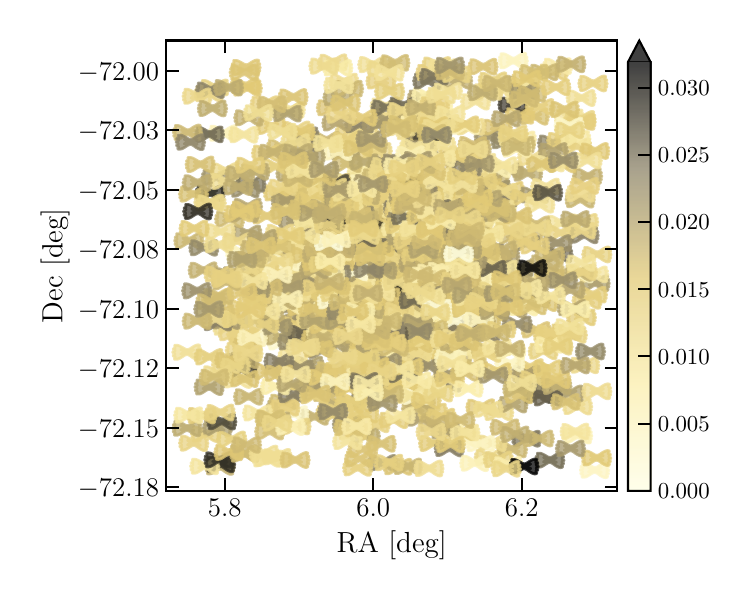}
    \caption{RGB stars from the inner regions of the globular cluster 47 Tucanae, sourced from \textit{Gaia} DR3 and visualized as a dish of farfalle with "cacio e pepe" sauce.}
    \label{fig:clusterpesto}
\end{figure}

In this Section, we present the upgraded geometry of the pasta markers and use the new colormaps and parmigiano features to demonstrate how astrophysical data visualization can achieve both heightened clarity and aesthetic distinction. Specifically, we showcase how conventional plotting methods, while foundational to the field, can be elevated into more memorable and intuitive representations, through the strategic application of culinary-inspired mapping.

\begin{figure}
    \centering
    \includegraphics[width=0.99\linewidth]{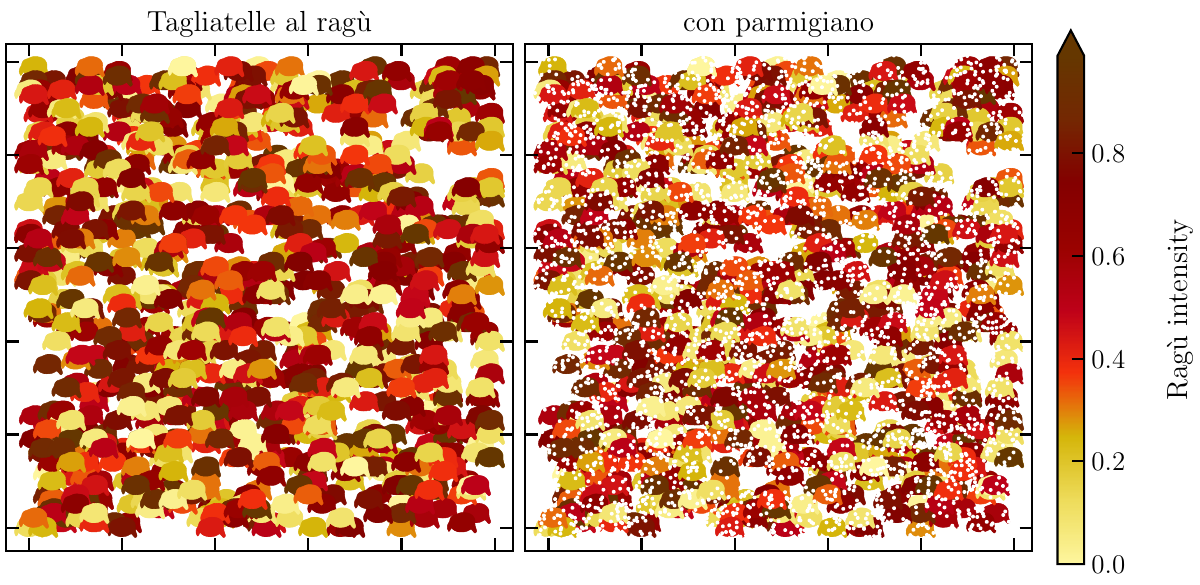}
    \caption{Comparison of the appearance of a randomly distributed cloud of points represented in tagliatelle with the ragu colormap without (left column) and with (right column) the addition of parmigiano.}
    \label{fig:parmigiano_sauce}
\end{figure}

In Figure \ref{fig:clusterpesto} we present a groundbreaking perspective of the Milky Way globular cluster 47 Tucanae (NGC 104) as farfalle with cacio e pepe sauce, to demonstrate the combination of using the upgraded versions of the markers with the new colormaps. We used only the brightest RGB stars in \textit{Gaia} DR3, with \texttt{phot\_g\_mean\_mag} $<15$ mag, and removed all stars with unreliable measurements in astrometry, such as parallax and proper motions.

\begin{figure}[!th]
    \centering
    \includegraphics[width=.50\textwidth]{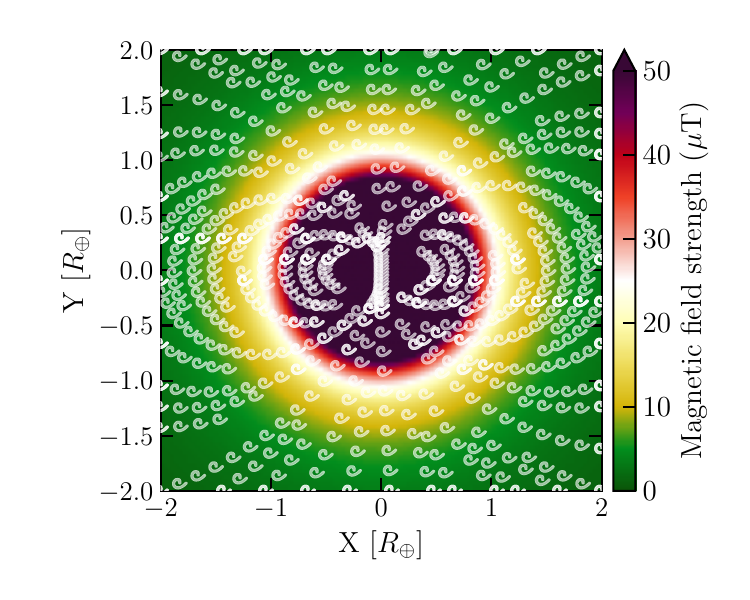}
    \caption{`Pasta alla norma' colormap of Earth and its surrounding region, color-coded by magnetic field strength, with gramigna pasta indicating the streams of the magnetic field lines.}
    \label{fig:earthmagnets}
\end{figure}

In Figure \ref{fig:parmigiano_sauce}, we show the comparison of a random distribution of tagliatelle with the ragu colormap with and without parmigiano. Depending on the taste of the user, each of them offers a unique and flavorful touch to the representation of data.

Figure \ref{fig:earthmagnets} is an approximation of the magnetic field of Earth, showing the strength of the magnetic field in terms of the `pasta alla norma' colormap, with magnetic field lines indicated by streams of gramigna pasta. The diameter of the Earth corresponds to the aubergine purple region of the colormap, with the surface of Earth indicated by the white ricotta cheese, and the surrounding area symbolized by yellow pasta and green basil.

\section{Conclusions}\label{sec:conclusions}
In this work, we have presented the second release of the \texttt{pastamarkers} package \cite{patamarkers_2024} that expands the available markers of \texttt{matplotlib} to include several pasta types. The pasta types we have chosen and implemented in the package are meant to enhance the visualization of astrophysical data, and ease the interpretation of complex data sets in the era of big surveys we are entering.

This release introduces two key improvements over the previous version. The first is the modification of the shape of the markers. Thanks to the refined geometry of the contours we have introduced, they are more similar to real pasta types. This significantly  enhances the clarity and aesthetics of visualization. The second one is the release of nine colormaps that are inspired by traditional pasta sauces. Their advantage is twofold: they facilitate the visualization of 3D data and they also boost the focus of the researchers compared to conventional colormaps. We claim this effect is due to an increased desire for lunch or snacks, depending on the time of the day the analysis is carried out, that leads to greater motivation in finishing work. Although promising, further research needs to be carried out to confirm or disprove this statement.

The code to use \texttt{pastamarkers} and corresponding colormaps is publicly available at this \href{https://github.com/LR-inaf/pasta-marker}{link}.

\begin{acknowledgements}

The authors thank the community for the positive feedback and interest for the first public release, that was a crucial motivation to work on the second one.

\end{acknowledgements}

\bibliographystyle{aa}
\bibliography{aanda.bib}

\end{document}